\documentclass[aps,floatfix,twocolumn,showpacs,prl]{revtex4-1}
\usepackage{graphicx,bm}
\usepackage{amsmath}
\usepackage{amssymb}
\usepackage{amsfonts}%
\usepackage{graphicx}
\usepackage{color}

\usepackage[T1]{fontenc}
\usepackage[latin2]{inputenc}
\usepackage{bm}

\newcommand{\bra}[1]{\langle #1|}
\newcommand{\ket}[1]{|#1\rangle}

\newcommand{\be}{\begin{equation}}
\newcommand{\ee}{\end{equation}}
\newcommand{\bea}{\begin{eqnarray}}
\newcommand{\eea}{\end{eqnarray}}

\newcommand{\Sz}{S_z}
\newcommand{\revision}[1]{{ #1}}

\begin{document}
\title{Limits of atomic entanglement by cavity-feedback : from weak to strong coupling}
\author{Krzysztof Paw\l owski}
\affiliation{Laboratoire Kastler Brossel,
Ecole Normale Sup\'erieure, UPMC and CNRS,
24 rue Lhomond, 75231 Paris Cedex 05, France}
\affiliation{Center for Theoretical Physics PAN, Al. Lotnik\'ow 32/46, 02-668
 Warsaw, Poland}
\author{J\'er\^ome Est\`eve}
\affiliation{Laboratoire Kastler Brossel,
Ecole Normale Sup\'erieure, UPMC and CNRS,
24 rue Lhomond, 75231 Paris Cedex 05, France}
\author{Jakob Reichel}
\affiliation{Laboratoire Kastler Brossel,
Ecole Normale Sup\'erieure, UPMC and CNRS,
24 rue Lhomond, 75231 Paris Cedex 05, France}
\author{Alice Sinatra}
\affiliation{Laboratoire Kastler Brossel,
Ecole Normale Sup\'erieure, UPMC and CNRS,
24 rue Lhomond, 75231 Paris Cedex 05, France}

\begin{abstract}
We theoretically investigate the entangled states of an atomic ensemble that can be obtained via cavity-feedback, varying the atom-light coupling from weak to strong, and including a systematic treatment of decoherence. 
In the strong coupling regime for small atomic ensembles, the system is driven by cavity losses into a long-lived, 
highly-entangled many-body state that we characterize analytically. In the weak coupling regime for  large ensembles, we find analytically the maximum spin squeezing that can be achieved by optimizing both the coupling and the atom number. This squeezing is fundamentally limited by spontaneous emission to a constant value, independent of the atom number.
\end{abstract}


\maketitle
Harnessing entanglement in many-body systems is of fundamental interest \cite{Amico:2008en} and is the key requirement for quantum enhanced technologies,
 in particular quantum metrology \cite{Giovannetti:2011jk}. 
In this respect, many efforts have been devoted to prepare entangled states in atomic ensembles because of their high degree of coherence 
and their potential for precision measurement. Spin squeezed states as well as 
number states have been produced following methods based either on coherent evolution 
in the presence of a non-linearity in the atomic field \cite{Oberthaler2010,Treutlein2010,vuletic2010EXP},
 or on quantum non-demolition measurement \cite{schleier-smith:2010vn,Thomson2014,Haas:2014kb}. 
Among methods of the first kind, cavity feedback \cite{vuletic2010EXP,SchleierSmith:2010eo} is one of the most promising: 
it has already allowed for the creation of highly squeezed states \cite{vuletic2010EXP} and the effective non-linearity introduced
 by the atom-cavity coupling can be easily switched off, making it very attractive for metrology applications. 

In this Letter, we analyze the entangled states that can be produced by cavity feedback in different coupling regimes from weak to strong, and derive the ultimate limits of the metrology gain, extending the optimization of squeezing to unexplored domains of parameters values. After optimization of both the coupling strength and the atom number, we find a maximum squeezing limit that depends only on the atomic structure. 
  
Cavity feedback relies on the dispersive interaction between one mode of an optical cavity and an ensemble of three level atoms, e.g. alkali atoms with a hyperfine splitting in the ground state (see Figure 1). The atom-cavity system is characterized by the atom-cavity coupling $g$, the cavity linewidth (HWHM) $\kappa$, the atomic detuning $\Delta$ and the spontaneous emission rate $\Gamma$ with ${\Delta\gg\Gamma}$. The dynamics of entanglement is governed by the two dimensionless quantities $C=g^2/(\kappa \Gamma)$ and ${\phi_0 = 2 g^2/(\kappa \Delta)}$. The cooperativity $C$ gives the ratio between the number of photons emitted in the cavity mode to spontaneously emitted photons as it can be seen by a Fermi golden rule argument \cite{AAMOP:2011}, and a large $C$ is favorable to entanglement because it minimizes the role of spontaneous emission. The parameter $\phi_0$ represents the cavity frequency shift, normalized to the cavity linewidth, when a single atom changes its hyperfine state. In the regime $\phi_0 \gg 1$, photons leaking from the cavity precisely measure the atom number difference between the two hyperfine states and therefore destroy coherence between them. One could expect that this may prevent the apparition of entanglement. However, this is not the case and we identify the condition to produce entanglement in this regime and characterize the produced states. They appear to have potential for metrology as signaled by their quantum Fisher information. In the regime $\phi_0 \ll 1$, spin coherence can be maintained and our calculations confirm that this regime is optimal for producing spin-squeezed states. One important result is that the maximum squeezing is limited by the ratio of the excited state linewidth to the hyperfine splitting that should be as small as possible. As $\phi_0/C=2\Gamma / \Delta$, the condition $\Gamma / \Delta \ll1$ allows to maintain $\phi_0$ small while maximizing $C$.
\begin{figure}[htb]
\centerline{\includegraphics[width=7.5cm,clip=]{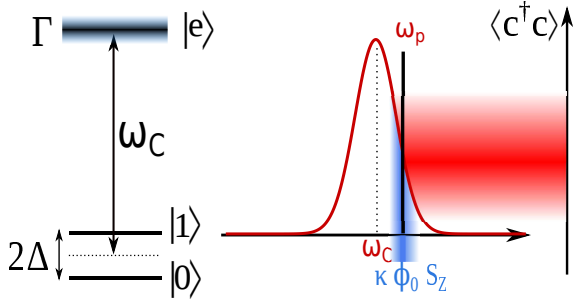}}
\caption{
Principle of the cavity feedback scheme. \revision{Left:} Three-level atoms are coupled to a cavity mode of frequency $\omega_c$. We suppose that the optical transitions between the two ground states and the excited state equally couple to the cavity mode with a coupling constant $g$. Thus, tuning the cavity as shown in the figure results in equal and opposite light-shifts for the two ground states. The width of the excited state is $\Gamma$. \revision{Right:} The evolution of the atomic state is induced by shining light onto the cavity at a frequency $\omega_p$ which is detuned from $\omega_c$ by $\delta$. 
The mean photon number in the cavity depends on $S_z$ the atom number difference between the two hyperfine states, via 
an atom-induced cavity detuning $\kappa \phi_0 S_z$ with $\phi_0$. When $\delta=\kappa$ and $\phi_0 \sqrt{N}$ 
is small compared to the cavity linewidth, the atom-induced fluctuation of the photon number is proportional to $\phi_0 S_z$
and the dynamics can be well approximated by an effective $\chi S_z^2$ model.
 \label{fig:Fig1}}
\end{figure}

We consider $N$ atoms, with two (hyperfine) ground states 
$|0\rangle$ and $|1\rangle$
equally coupled with a constant $g$ and opposite detunings $\pm \Delta$ to an excited manifold $|e\rangle$ by a single cavity mode (see Fig.\ref{fig:Fig1}a). 
We introduce the collective spin operators $S_{x}+iS_y=\sum_{i=1}^N |1\rangle \langle0|_i$, $S_{z}=\frac{1}{2}\sum_{i=1}^N |1\rangle \langle1|_i-|0\rangle \langle0|_i$ obtained by summing the effective spin 1/2 operators for each atom.
The initial atomic state is a coherent spin state, each atom being in an even superposition of $\ket{0}$ and $\ket{1}$.
A single off-resonant cavity photon shifts the energies of these levels in opposite directions by an amount 
$\frac{g^2}{\Delta}$. 
Through these opposite light shifts, the energy difference between levels $\ket{0}$ and $\ket{1}$ depends on the cavity photon number $c^\dagger c$, that depends on its turn on the population difference $S_z$ as the atoms change the index of refraction in the cavity. Spin squeezing in this scheme occurs in the following way:
the atomic quantum noise in $S_z$, induces fluctuations of the cavity field intensity 
(as shown in Fig.\ref{fig:Fig1}), which during the evolution are imprinted into the phases of each atom,
thus correlating $S_y$ with the population imbalance $S_z$ \footnote{Spin squeezing is indeed quantum correlations between the two transverse components $S_y$ and $S_z$ of the collective spin}.
Assuming low saturation of the optical transition $g^2\langle c^\dagger c\rangle/\Delta^2\ll1$, we eliminate the excited manifold $|e\rangle$ and describe each atom within the
$|0\rangle$-$|1\rangle$ subspace. The unitary evolution is governed by
	\begin{equation}
	H_{0}/\hbar = (\delta+\kappa \phi_0 S_z) c^\dagger c 
	+ i \eta (c^\dagger-c) 
	\label{eq:H0}
	\end{equation}
where $c$ annihilates a photon of the cavity mode, $\eta$ is the cavity pumping rate, $\delta=(\omega_c-\omega_p)$ is the empty cavity detuning, 
and we already introduced $\phi_0 = 2 g^2/(\kappa \Delta)$ that is also the single-photon atomic light shift properly normalized.
Cavity losses and spontaneous emission 
\revision{including the possibility to scattering outside the $\ket{0}-\ket{1}$ subspace}
are described by jump operators: $d_c=\sqrt{2\kappa} c$ and \cite{Cohen1990,Ozeri2010,Gerbier2010}
\begin{eqnarray}
d_{i,{\rm el}}&=&\sqrt{\frac{\Gamma_{\rm Ray}}{2}}(|1\rangle \langle1|-|0\rangle \langle0|)_i c \:;\:  
\frac{\Gamma_{\rm Ray}}{2}=\frac{\Gamma \phi_0}{\Delta} a_{\sigma \sigma} \quad \\
d_{i,\sigma'\sigma} &=& \sqrt{\Gamma_{\rm Ram}} |\sigma' \rangle \langle \sigma|_i c \:;\:  
\Gamma_{\rm Ram}=\frac{\Gamma \phi_0}{\Delta} \frac{|a_{\sigma'\sigma}|^2}{a_{\sigma \sigma}} \\
d_{i,X\sigma} &=& \sqrt{\Gamma_{\rm X}} |X \rangle \langle \sigma|_i c  \:;\:  
\Gamma_{X}=\frac{\Gamma \phi_0}{\Delta} \frac{\sum_{X\neq 0,1} |a_{X\sigma}|^2}{a_{\sigma \sigma}} 
\label{eq:jumps}
\end{eqnarray}
here $\sigma,\sigma'=0,1$ and $X\neq 0,1$ label the internal state, $i$ labels the atom
and $a_{\sigma' \sigma}$ are amplitudes  that depend on the atomic structure and field polarization
\cite{suppl}. The operator $d_{i,0}$, $d_{i,\sigma'\sigma}$, $d_{i,X\sigma}$ refer to Rayleigh and Raman processes for the atom $i$. For any given eigenstate of the atomic operator $S_z$ with eigenvalue $m \in [-N/2,N/2]$, the cavity field reaches in a time $1/\kappa$ a steady state that is a coherent state of amplitude $\alpha(m)$
	\begin{equation}
	\alpha(m) =   \frac{\eta}{  \kappa_{\rm eff} + i (\delta +\kappa \phi_0 m) }
	\label{eq:alpha_m} 
	\end{equation}
where $\kappa_{\rm eff}$ is the cavity linewidth in presence of the atoms 
	\begin{equation}
	\kappa_{\rm eff} =  \kappa \left( 1 +  \frac{N}{4} 
	(\Gamma_{\rm Ray}+\Gamma_{\rm Ram}+\Gamma_{\rm X}) \right) \stackrel{N\ll \frac{ \Delta}{\Gamma \phi_0}}{\simeq} \kappa
	\label{eq:keff}
	\end{equation}
As we are interested in times $t\gg \kappa^{-1}$, we shall neglect the transient effects and assume that the cavity field
is in steady state from the beginning of the evolution \footnote{A sufficient condition for this approximation to be accurate is that average photon number in the cavity is small $(\eta/\kappa)^2\ll 1$, which guarantees that the probability of having a cavity jump during the transient time is small.}. 
In alkali atoms with a small hyperfine splitting in the excited state, choosing the states $m_F=0$ for $|0\rangle$ and $|1\rangle$, $\pi$-polarized light and a detuning close to half of the hyperfine energy splitting, one finds opposite light-shifts for the two states and no Raman processes coupling $|0\rangle$ and $|1\rangle$.
In this first example we therefore restrict our analysis to Rayleigh processes that commute with $S_z$. Raman spin-flipping processes and scattering to other states will be discussed later on in particular in relation to spin squeezing.
Under these conditions, with $a_{11}=a_{22}$ and $a_{\sigma \sigma'}=0$ for $\sigma \neq \sigma'$, one can calculate the atomic density matrix $\rho$, expressed in the tensor product basis $|\vec{\epsilon} \, \rangle = | \epsilon_1,  \epsilon_2, \ldots,  \epsilon_N \rangle $ where $\epsilon_i=0,1$ refers to the internal state of the $i$-th atom
	\begin{eqnarray}
	\langle \vec{\epsilon}_1 | \rho | \vec{\epsilon}_2 \rangle &=& \frac{1}{2^N} 
	\langle \alpha(m_{\vec{\epsilon}_1}) | \alpha(m_{\vec{\epsilon}_2}) \rangle^{1+2\kappa t + (N-\|  \vec{\epsilon}_1 - \vec{\epsilon}_2 \|){\Gamma}_{\rm Ray}t} \nonumber \\
	&\times& e^{ i \kappa t \left[ | \alpha(m_{\vec{\epsilon}_1})|^2\left( \delta/\kappa+ \phi_0 m_{\vec{\epsilon}_1} \right) 
	- | \alpha(m_{\vec{\epsilon}_2})|^2\left(  \delta/\kappa+ \phi_0 m_{\vec{\epsilon}_2} \right) \right]  } \nonumber \\
	&\times& e^{-\left( | \alpha(m_{\vec{\epsilon}_1})|^2 + | \alpha(m_{\vec{\epsilon}_2})|^2\right) \| \vec{\epsilon}_1 - \vec{\epsilon}_2 \|	\frac{{\Gamma}_{\rm Ray}t}{2} } \,,
	\label{eq:rhomatrix}
	\end{eqnarray}
where $\|\vec{\epsilon}\,\| \equiv \sum_{i=1,N} |\epsilon_i|$ and $m_{\vec{\epsilon}}=\|\vec{\epsilon}\,\|-N/2$. 
The first line in (\ref{eq:rhomatrix}) represents decoherence due to loss of photons (through cavity losses and  spontaneous emission) that are entangled with the atoms. The second line represents the unitary evolution, whereas the third line is a second contribution of spontaneous emission that tends to kill all the off-diagonal elements of the density matrix by projecting single atoms into $\ket{0}$ or $\ket{1}$.
	\begin{figure}[htb]
	\centerline{\includegraphics[width=7cm,clip=]{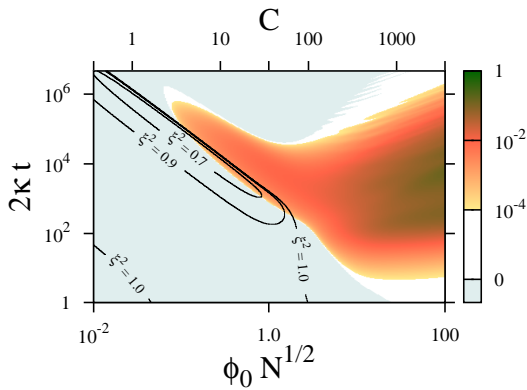}}
	\caption{(Color online) Map of entangled states at different times and values of the phase shift $\phi_0$. The purity change is shown in color and the 
	isolines of the squeezing parameter are solid black lines. Parameters : $\delta=\kappa$, $N=50$, $(\eta/\kappa)^2=10^{-2}$, $a_{\rm 11}=a_{\rm 22}=1.0$, $a_{\sigma \sigma'}=0$ for $\sigma \neq \sigma'$, $\Delta/\Gamma=500$.
	\label{fig:Fig2}}
	\end{figure}
To explore the apparition of entanglement in the evolution starting from a coherent spin state $|\psi(0)\rangle = [(|0\rangle+|1\rangle)/\sqrt{2}]^{N}$, we calculate the change in the system purity after tracing out one atom \cite{Cerf1999, Rossignoli2002} that we note ${\rm PC}$ ({\it purity change}):
	\begin{equation}
	{\rm PC} \equiv {\rm Tr}_{1,2,\ldots ,N}[\rho^2]-{\rm Tr}_{2,\ldots,N}[({\rm Tr}_1 \rho)^2]
	\label{eq:PC}
	\end{equation}
If all the atoms are correlated, tracing one of them can strongly influence the purity.
Indeed, one can show that ${\rm PC}>0$ implies that the state is not separable.
${\rm PC}$'s maximum value is 1/2 obtained for a 
Schr\"odinger cat state. In a purely hamiltonian model $H=\chi S_z^2$, one has ${\rm PC}=\frac{1}{2}[1-(\cos \chi t)^{2(N-1)}]$ \cite{suppl}.
The main advantage of the quantity \eqref{eq:PC}, is that it only requires the calculation of a trace and it can be computed even for relatively large atom numbers despite the large size of the Hilbert space.
In Fig. \ref{fig:Fig2} we show PC, as a function of time and of $\phi_0 \sqrt{N}$ that is the cavity detuning induced by the quantum fluctuations of $S_z$ (as $\Delta S_z=\sqrt{N}/2$).
On the same plot we show the isolines for the spin squeezing parameter \cite{winenland1992}
	\begin{equation}
	\xi^2=\frac{N\Delta^2 S_\perp}{|\langle \vec{S} \rangle|^2}
	\end{equation}
where $\Delta^2 S_\perp$ is the minimal variance of the collective spin orthogonally to the mean spin direction and 
$|\langle \vec{S} \rangle|$ is the mean spin length. Spin squeezed states $\xi^2<1$ appear for coupling values such that 
$\phi_0 \sqrt{N} < 1$, this conclusion illustrated for 50 atoms in Fig.\ref{fig:Fig1}, holds for much larger atom numbers (see Fig.\ref{fig:Fig4}).

\begin{figure}[htb]
		\centerline{\includegraphics[width=8.3cm,clip=]{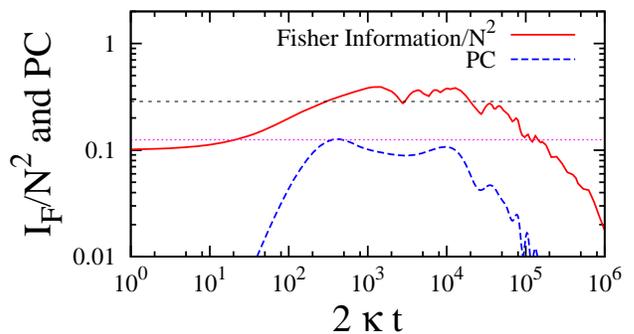}}
		\caption{(Color online) Purity-change (blue dashed line) and Fisher information (red solid line) as a function of time for 
		$\phi_0=14$, $N=10$ in presence of Rayleigh scattering. The horizontal lines give the analytical predictions for 
		$I_F$ and $PC$ obtained from the state (\ref{eq:werner}).
	      Other parameters are as in Fig.\ref{fig:Fig2}.
		\label{fig:Fig3}}
	\end{figure}

\begin{figure}[htb]
		\centerline{\includegraphics[width=8.3cm,clip=]{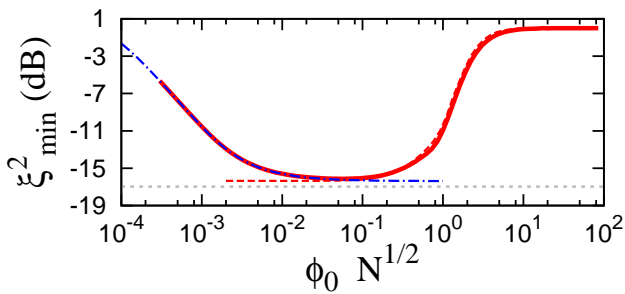}}
		\caption{(Color online) Spin squeezing optimized over time as a function of $\phi_0 \sqrt{N}$ in a realistic configuration for 
		${}^{87}$Rb (see text) with $N=10^5$,  $\Delta/\Gamma=563.39$ and $(\eta/\kappa)^2=10^{-2}$ and $\delta=\kappa$.
		Red solid line: full model (\ref{eq:rhomatrix}) with cavity losses and Rayleigh jumps ($a_{11}=a_{22}=0.702$). 
		Red dashed line: full model without spontaneous emission. 
		Blue dash-dotted line: effective model in the regime $\phi_0 \sqrt{N} \ll 1$ and $\kappa_{\rm eff}\simeq\kappa$ with $a_{11}=a_{22}=0.702$, $a_{0,1}=a_{1,0}=0$, and $a_{\rm X,1}=a_{\rm X,0}=0.497$. \revision{
		Horizontal gray dotted line: $\xi_{\rm min}^2$ in the large $N$ limit (\ref{eq:xi2-below-Nc}).}
		\label{fig:Fig4}}
	\end{figure}

{\it Strong coupling regime -}
The purity change ${\rm PC}$ in Fig.\ref{fig:Fig2} detects a larger and larger region of entangled states as $\phi_0 \sqrt{N}$ increases, even after very short evolution times. 
To understand their nature, let us first consider the case without spontaneous emission.
In this case decoherence is only due to cavity losses 
and one can use the Fock basis $\ket{m}$ to express the density matrix whose off-diagonal elements
decay as $\langle \alpha(m) | \alpha(m^{\prime}) \rangle^{1+2\kappa t }$. If $\phi_0 \gg 1$,
any state with $m\neq0$ shifts the cavity out of resonance so that the cavity is practically dark,
while the only distinguished state is the twin-Fock state $m=0$ which does not detune the cavity. Coherences between this state and all the others vanish in a time
$t_0\simeq\frac{1}{\kappa |\alpha(m=0)|^2}=2 \kappa/\eta^2$ after which
the initial coherent spin state $\ket{\psi} = \sqrt{p_{0}}  \ket{m=0} + \sqrt{(1-p_{0})}\ket{\psi^{\perp}}$
\footnote{$p_m= \sqrt{\frac{1}{2^N} \binom{N}{m+\frac{N}{2}}}$} is mapped onto the mixture:
	\begin{equation}
	 \rho= p_{0}  \ket{m=0}\bra{m=0} + (1-p_{0}) \ket{\psi^{\perp}}\bra{\psi^{\perp}}
	 \label{eq:werner}
	\end{equation}
The subspace $\ket{\psi^{\perp}}\bra{\psi^{\perp}}$ of states $m\neq0$ is preserved, both from decoherence and unitary evolution, for a time $t_1 \simeq \frac{1}{\kappa |\alpha(m=1)|^2}=(\kappa/\eta^2) \phi_0^2/4$ by which the cavity starts to distinguish the next Fock state. At longer times more and more coherences between different Fock states are killed until a complete mixture is reached. Note that $t_1/t_0=\phi_0^2/8\gg1$ for strong coupling.
Interestingly, the long-lived state (\ref{eq:werner}) with partially removed coherences is highly entangled, its Fisher information scaling as $I_F=2\sqrt{2/\pi} N^{3/2}$ for large $N$. 
In Fig.\ref{fig:Fig3} we show the purity-change and the Fisher information as a function of time, for $\phi_0=14$ and $N=10$ atoms. Rayleigh spontaneous emission and cavity losses are included. Fisher information and PC reach those 
of the state (\ref{eq:werner}) around the time $\kappa t =250\simeq 1.25\kappa t_0$ and stay close to these values until 
$\kappa t=2\times10^4\simeq 2\kappa t_1$ indicating that our picture still holds in presence of spontaneous emission for small atomic ensembles. \footnote{For PC, a similar conclusion holds even for $N=50$.}
From the last row of Eq. \eqref{eq:rhomatrix} we see however that there are density matrix terms that are
very sensitive to spontaneous emission and decay in a timescale $1/N\Gamma $, suggesting that 
the non-Gaussian state \eqref{eq:werner} is probably limited to small numbers of atoms.

{\it Spin-squeezed states -}
We now concentrate on spin squeezing and large atom numbers. 
In Fig.\ref{fig:Fig4} we show 
the squeezing parameter optimized over time $\xi^2_{\rm min}$, as a function of $\phi_0 \sqrt{N}$ for $N=10^5$, in a realistic configuration for ${}^{87}$Rb where we 
choose the clock transition $|F=1,m_F=0\rangle-|F=2,m_F=0\rangle$, $\pi$-polarized light on D2 line
and a detuning close to half of the hyperfine energy splitting so that there are symmetric couplings $a_{11}=a_{22}$ and no Raman processes $a_{12}=a_{21}=0$ \cite{suppl}.
The red solid curve, calculated from (\ref{eq:rhomatrix}) includes cavity losses and Rayleigh spontaneous emission while the red dashed curve includes only cavity losses. 
\revision{We see that spontaneous emission is here important only for small values of $\phi_0 \sqrt{N}$. The blue dash-dotted curve is an effective $H=\chi S_z^2$ model which we derive for $\phi_0 \sqrt{N} \ll1$ and $\kappa_{\rm eff}\simeq \kappa$, in which we can include all the loss processes and that we can solve analytically \cite{suppl}. For large enough coupling $N^{-1/10} \ll \phi_0 \sqrt{N}$
this models predicts a best squeezing limited by cavity losses as found in \cite{SchleierSmith:2010eo,SchleierSmith:2012}
\begin{equation}
		\xi^2_{\rm min} = \frac{5}{6}  \left( 3\right)^{4/5}  N^{-2/5} \quad;\quad
		 t_{\rm min} = \frac{2}{\eta^2\phi_0^2} \left( 3\right)^{1/5} N^{-3/5}.
		\label{eq:xi2-below-Nc}
\end{equation}
On the other hand, if $\phi_0 \sqrt{N}\geq1$ the squeezing is lost in the full model because of non linearities 
that are not in the effective model that is here out of its limit of validity.
Nonlinear effects come into play when the cavity detuning to due quantum fluctuations $\kappa \phi_0 \sqrt{N}$ exceeds the effective cavity linewidth $\kappa_{\rm eff}$ (the blue region in Fig. \ref{fig:Fig1} exceeds size of the fringe).
As we show now, the equality condition between theses two quantities allows to introduce a critical atom number 
$N_c$ that distinguishes between two different regimes for spin squeezing. In each of these regimes, $N<N_c$ and $N>N_c$, an appropriate effective $H=\chi S_z^2$ model can be derived in some parameter range.
Using (\ref{eq:keff}), the condition $\kappa \phi_0 \sqrt{N} \leq \kappa_{\rm eff}$ can be written as
	\begin{equation}
	\phi_0 \sqrt{N} \left( 1 - \sqrt{\frac{N}{N_c}}\right) \leq 1 \:; \:\: N_c\equiv  \left( 
	\frac{4\Delta/\Gamma}{a_{\rm Ray}+a_{\rm Ram}+a_{\rm X}} \right)^2
	\label{eq:condition}
	\end{equation}
where we have introduced $a_{\rm Ray}=2a_{\sigma,\sigma}$, 	$a_{\rm Ram}=|a_{\sigma',\sigma}|^2/a_{\sigma,\sigma}$
and $a_{\rm X}=\sum_{X\neq 0,1} |a_{X\sigma}|^2/a_{\sigma \sigma}$.
$N_c$ corresponds to such number of atoms that photon losses due to atom-scattering equal those due mirror transmission.

{\it For $N\ll N_c$} which is the situation of Fig.\ref{fig:Fig4}, the condition (\ref{eq:condition}) gives $\phi_0 \sqrt{N} \leq1$.
For $\phi_0\sqrt{N}>1$ the detuning induced by the quantum noise becomes larger than the cavity linewidth giving rise to nonlinear effects destroying squeezing.

{\it For $N\gg N_c$} we always have $\phi_0 \sqrt{N} \ll \kappa_{\rm eff}$.
In this regime dominated by absorption, the cavity linewidth increases linearly with the atom number $\kappa_{\rm eff}/\kappa \approx  N\phi_0/\sqrt{N_c}$, faster than the atoms induced cavity detuning $\propto\phi_0\sqrt{N}$.
By deriving a second effective $H=\chi S_z^2$ model, for $\kappa_{\rm eff} \gg \kappa$ and $N \gg N_c$ for cavity losses and Rayleigh jumps \cite{suppl},
we find that (i) the best squeezing becomes independent of $\phi_0$ for large $\phi_0\sqrt{N}$ and, most importantly,
(ii) the best squeezing has a non-zero limit for $N\to \infty$
	\begin{equation}
		\xi^2_{\rm min} \stackrel{N\to \infty}{\to} e\, \left(\frac{2 a_{11} \Gamma }{\Delta}\right)^{2}
	\label{eq:xi2best-case2}
	\end{equation}
We show the onset of this new regime as $N$ is increased in Fig.\ref{fig:Fig5} for $\Delta/\Gamma=10$.
\begin{figure}[htb]
\includegraphics[width=7.5cm,clip=]{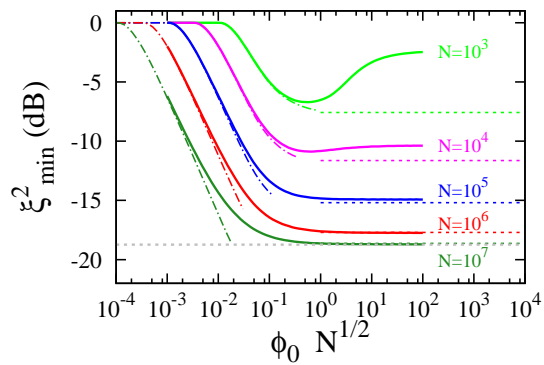}
\caption{
Spin squeezing optimized over time as a function of $\phi_0 \sqrt{N}$, for  $\Delta/\Gamma=10$ and $(\eta/\kappa)^2=10^{-2}$.
Solid lines: full model (\ref{eq:rhomatrix}) with cavity losses and Rayleigh scattering with $a_{11} =a_{22}=0.702$. 
Dot-dashed lines: analytical results in the regime $\kappa_{\rm eff}\simeq\kappa$ and  $\phi_0 \sqrt{N} \ll 1$.
Dotted lines: analytical results in the regime  $\kappa_{\rm eff}\gg \kappa$  and  $\phi_0 \sqrt{N} \gg 1$. Horizontal gray dotted line: $\xi_{\rm min}^2$ in the large $N$ limit (\ref{eq:xi2best-case2}).
From top to bottom: $N= 10^3,\;10^4,\;10^5,\;10^6,\;10^7$.
\label{fig:Fig5}}
\end{figure}
 }
 
{\it Conclusions} We predicts that highly-entangled many-body states driven by cavity losses can be prepared by cavity feedback in the strong coupling regime for small samples, and a very large amount of spin-squeezing is reachable for large atom numbers in the weak coupling regime. The spin-squeezing limit we find (\ref{eq:xi2best-case2}) is a very small
value for alkali atoms suggesting that there is still room for improvement in the experimental achievements.
\begin{acknowledgments}
The authors would like
We are grateful to R. Kohlhaas and Y. Castin for discussions. 
The work was supported by the European QIBEC project, by C'Nano Ile de France CQMet project and by the (Polish) National Science Center Grant No. DEC-2012/04/A/ST2/0009.
\end{acknowledgments}

\end{document}